\documentclass[12pt]{article}

\usepackage{amsmath,amssymb}

\newcommand{\beq}{\begin{eqnarray}}
\newcommand{\eeq}{\end{eqnarray}}
\newcommand{\rf}[1]{(\ref{#1})}

\begin{document}

\thispagestyle{empty}

\begin{center}

%\hspace{8cm}

\begin{flushright}
NSF-KITP-05-04\\
BCCUNY-HEP/05-01
\end{flushright}

\vspace{15pt}
{\large \bf LATTICE
$({\rm QCD})_{2+1} $    }

\vspace{20pt}

{\bf Peter Orland}$^{\rm a.}$$^{\rm b.}$$^{\rm c.}$

\vspace{8pt}

\begin{flushleft}
a. Kavli Institute for Theoretical Physics, The University of California, Santa Barbara, CA 
93106, U.S.A.
 \end{flushleft}
 
\begin{flushleft}
b. Physics Program, The Graduate School and University Center,
The City University of New York, 365 Fifth Avenue,
New York, NY 10016, U.S.A.
\end{flushleft}

\begin{flushleft}
c. Department of Natural Sciences, Baruch College, The 
City University of New York, 17 Lexington Avenue, New 
York, NY 10010, U.S.A., orland@gursey.baruch.cuny.edu 
\end{flushleft}

\vspace{40pt}

{\bf Abstract}
\end{center}

\noindent 
We consider a
 $(2+1)$-dimensional SU($N$) lattice gauge theory in an axial gauge with the link field
 $U_{1}$ set equal to one. The term
 in the Hamiltonian containing the square of the electric field in the $1$-direction 
 is non-local. Despite this
 non-locality, we show
 that weak-coupling perturbation theory in this term gives a finite vacuum-energy 
 density to second order, and suggest that this property holds to all orders. Heavy
 quarks are confined, the spectrum is gapped, and the space-like Wilson loop has area decay.

\hfill

\newpage

\setcounter{page}{1}

\section{Introduction}
\setcounter{equation}{0}
\renewcommand{\theequation}{1.\arabic{equation}}

The central problem of QCD is confinement. It is not enough to prove that lattice
gauge theories have a confining phase - which is evident from strong-coupling
expansions. It is necessary to see that the color is confined at arbitrarily weak lattice
coupling.

We find here that $(2+1)$-dimensional lattice gauge theories 
confine for any dimensionless bare coupling. The technique used is a {\em weak-coupling}
expansion in an anisotropic lattice gauge theory in an axial gauge.  This is not
the standard expansion utilizing Feynman diagrams. Though the coupling
constant of a $(2+1)$-dimensional gauge theory 
is not infinitely renormalized in $(2+1)$ dimensions, the dimensionless bare
coupling on the lattice must vanish in the continuum limit. This is why a weak-coupling analysis is
useful, even for this case. The dependence we find of the string tension and the mass gap
on the coupling
constant does not agree with  conventional wisdom  - our results for these 
physical quantities don't behave as anticipated, as the lattice spacing is taken to
zero - but
they are not zero.

The first analytic demonstration of confinement of heavy sources 
in $(2+1)$-dimensional
gauge theories was given by Polyakov for lattice compact QED \cite{polyakov1}, and later for the
Georgi-Glashow model \cite{polyakov2}.  The latter model is interesting
in that color charges disappear completely from the spectrum.  This is, however, different
from the sort of confinement we expect for QCD, in that matter fields
play an important role. Feynman argued that $(2+1)$-dimensional
QCD is confining \cite{feynman}.  Unfortunately, the orbit-space distance estimates
in Feynman's 
paper are incorrect. Nonetheless, his basic claim, that the diameter of gauge-orbit space
of $(2+1)$-dimensional SU($N$) Yang-Mills theory for small magnetic energy is finite, appears
to be correct  \cite{orl-sem}.  New nonperturbative methods which do not
require a lattice have been derived by Karabali, Kim and Nair
\cite{kar-nair} (one of their formulations of the Hamiltonian has been obtained a
different way in reference \cite{orl-gauge-inv}).

There is one special set of assumptions we use to derive 
our results; the $(1+1)$-dimensional
non-Abelian nonlinear sigma models (without topological terms)
have a mass gap, exponentially decaying correlation
functions, and their vacuum expectation values of local operators exhibit clustering 
\cite{sigma}, \cite{pol-wieg}, \cite{abda-wieg}.  Though no rigorous proof of these
properties exists, we think that the evidence in their favor is overwhelming.

Our basic strategy is to write the lattice 
version of the  Hamiltonian as the sum of two terms, namely
\begin{eqnarray}
H_{0}=\int d^{2}x  \; \left(\frac{e^{2}}{2} {\rm Tr} \,{\mathcal E}_{2}^{2}+\frac{1}{2e^{2}}{\rm Tr}\,B^{2}\right)\;, \nonumber
\end{eqnarray}
and 
\begin{eqnarray}
\frac{e^{2}}{2}V=\frac{e^{2}}{2} \int d^{2}x \;{\rm Tr}\, {\mathcal E}_{1}^{2} \;, \nonumber
\end{eqnarray}
where ${\mathcal E}_{j}$ are the components of the electric field conjugate to the gauge field
$A_{j}$, $[{\mathcal E}_{j}(x),A_{k}(y)]={\rm i}\delta_{jk}\delta^{2}(x-y)$ and $B={\rm i}[\partial_{1}-{\rm i}A_{1},\partial_{2}-{\rm i}A_{2}]$ is the single space
component of the magnetic field. We then pick the gauge $A_{1}=0$. When this is done on the
lattice, $H_{0}$ is a set of
decoupled chiral SU($N$)$\times$SU($N$) nonlinear sigma models for which the $S$-matrix and
the spectrum are
known. The quantity $V$ is non-local, but we show that perturbation theory in this term is sensible
to second order.  The vacuum state in this perturbation series confines fundamental color
charges. Our splitting of the Hamiltonian is not explicitly rotationally 
invariant, but if the method works to 
all orders of perturbation theory, rotational invariance should be restored.  

Let us review 
axial gauges in the continuum.  If the 
SU($N$)-Lie-algebra-valued gauge field $A_{1}$ is set to zero, then Gauss's 
law may be integrated to obtain
\begin{eqnarray}
{\mathcal E}_{1}(x) 
&\!=\!&
-\int^{x^{1}} \!dy^{1} \;\sum_{j=2}^{d-1}
[\partial_{j}-iA_{j}(y^{1},x^{2},\dots x^{d-1}), {\mathcal E}_{j}(y^{1},x^{2},\dots,x^{d-1})] \nonumber \\
&\!=\!&-\int^{x^{1}} \! dy^{1}\;
 { D}_{\perp}(y^{1},x^{\perp}) \cdot
{\mathcal E}_{\perp}(y^{1}, x^{\perp})
\;, \label{continuum-electric}
\end{eqnarray}
where the dimension of space is $d-1$, $x^{\perp}=(x^{2},\dots,x^{d-1})$, and
${ D}_{\perp}$ are the
last $d-2$ components of  the covariant derivative
in the adjoint representation 
$({ D}_{2},\dots,{ D}_{d-1})$.  The term in
the Hamiltonian
\begin{eqnarray}
\frac{e^{2}}{2}V=\int d^{d-1} x \;\frac{e^{2}}{2} {\rm Tr}\,  {\mathcal E}_{1}^{2} \label{continuum-electric-energy}
\end{eqnarray}
must have a vacuum expectation value proportional to the volume of $(d-1)$-dimensional
space, if the theory is to be sensible.  As discussed by Mandelstam 
\cite{mandelstam}  this means that the quantity
\begin{eqnarray}
K(y^{1},z^{1},x^{\perp} )\! =\!
\left<0 \vert  {\rm Tr}\;{ D}_{\perp}(y^{1},x^{\perp}) \!\cdot \!
{\mathcal E}_{\perp}(y^{1}, x^{\perp})\;
{ D}_{\perp}(z^{1},x^{\perp}) \!\cdot \! {\mathcal E}_{\perp}(z^{1},x^{\perp})
\vert 0 \right>  
\nonumber
\end{eqnarray}
must have the property that $\int dy^{1}dz^{1}K(y^{1},z^{1},x^{\perp})$ does not diverge with
the spatial volume. One might think that if 
$K( y^{1} ,z^{1},  x^{\perp} )$
falls off sufficiently fast with $\vert y^{1}-z^{1}\vert$, the problem can be ameliorated. Rapid
fall-off of $K$, however,  is not enough. Even if the fall-off is exponential, the result may diverge as
$(L^{1})^{2}$ where $L^{1}$ is the range of $x^{1}$. Mandelstam recognized that the residual
gauge invariance, remaining after solving for ${\mathcal E}_{1}$ in (\ref{continuum-electric}), namely
\begin{eqnarray}
\int  dx^{1} \;{ D}_{\perp}\cdot {\mathcal E}_{\perp} \; \Psi =0\;, \label{finiteness}
\end{eqnarray}
must also be satisfied by the vacuum. Without both (\ref{finiteness}) and
and sufficiently rapid decay of $K(x^{1},y^{1}, x^{\perp})$,
any conjecture
for the vacuum may have an unacceptable infrared-divergent energy, coming
from (\ref{continuum-electric-energy}). Fortunately, we 
find that in our perturbation scheme, both 
the unperturbed vacuum energy and the first two corrections in
our weak-coupling expansion
obey the lattice versions of both the rapid-decay criterion and (\ref{finiteness}).

\section{The lattice gauge Hamiltonian}
\setcounter{equation}{0}
\renewcommand{\theequation}{2.\arabic{equation}}

The purpose of this section is to establish our definitions and
conventions. It is not an introduction to the 
Hamiltonian SU($N$) gauge 
theory. Such introductions can be found in the review 
article by Kogut and 
in the book by Creutz \cite{kogut}.

Consider a lattice of sites $x$ of size
$L^{1}\times L^{2}$, with
sites $x$ whose coordinates are $x^{1}$ and $x^{2}$. We require that $x^{1}/a$ and $x^{2}/a$
are integers, where
$a$ is the lattice spacing. There
are $2$ space directions, labeled
$j=1,2$. Each link is a pair $x$, $j$, and joins the site 
$x$ to $x+{\hat j}a$, where
$\hat j$ is a unit vector in the $j^{\rm th}$ direction.

We introduce basis vectors or generators 
$t_{\alpha}$, $\alpha=1,\dots, {N}^{2}-1$, of 
the Lie algebra
of SU($N$). Sometimes we use Roman letters for
the index, e.g. we may write $t_{b}$ rather than $t_{\alpha}$ (the
purpose of using different alphabets is to distinguish between
coordinate indices on the SU($N$) manifold and tangent-space
vectors). The
generators are defined to be orthonormal, so that 
Tr$\;t_{\alpha}t_{\beta}
=\delta_{\alpha \beta}$. The structure
coefficients of the Lie algebra, $f_{\alpha \beta}^{\gamma}$, 
$\alpha, \beta, \gamma =1,\dots, {N}^{2}-1$, are, as usual, the 
complex numbers defined by
$[t_{\alpha}, t_{\beta} ]=i f_{\alpha \beta}^{\gamma} t_{\gamma}$. The
identity matrix will be denoted by 
${\rm 1}\!\!\,{\rm l}$.

The Hamiltonian lattice gauge theory is usually formulated in
temporal gauge $A_{0}=0$. The basic degrees of freedom, before 
any further 
gauge fixing, are elements of the group SU($N$) in the 
fundamental ($N\times N$)-dimensional 
matrix representation $U_{j}(x)\in$ SU($N$) at each link
$x$, $j$. In addition, there are the ${N}^{2}-1$ electric-field 
operators at each link
$l_{j}(x)_{b}$, $b=1,\dots, {N}^{2}-1$. The electric-field 
operators
are self-adjoint by construction. The 
commutation relations on the
lattice are
\begin{eqnarray}
[l_{j}(x)_{b} , l_{k}(y)_{c} ]=
i\delta_{x\;y}\delta_{j\;k} \;f_{b c}^{d}
\;l_{j}(x)_{d} \;, \nonumber 
\end{eqnarray}
\begin{eqnarray}
[l_{j}(x)_{b}, U_{k}(y)]        =
-\delta_{x\;y}\delta_{j\;k}\; t_{b}\;U_{j}(x)\;,
\label{loccommrel}
\end{eqnarray}
all others zero. In 
the Schr\"{o}dinger
representation, with the components of $U_{j}(x)$ taken to be 
c-numbers, 
the latter of \rf{loccommrel} becomes
\begin{eqnarray}
l_{j}(x)_{b} U_{k}(y)=-\delta_{x\;y}\delta_{j\;k} 
\;t_{b}\;U_{j}(x)\;,
\nonumber
\end{eqnarray}

The lattice Hamiltonian is
\begin{eqnarray}
H= \sum_{x } 
\sum_{j=1}^{2} \sum_{b=1}^{{N}^{2}-1}
\;\frac{g_{0}^{2}}{2a}\left[ \, l_{j}(x)_{b} \,\right]^{2}-
\sum_{x} 
\frac{1}{4g_{0}^{2}a}\;
\left[ {\rm Tr}\; U_{1\,2}(x) +{\rm Tr}\;U_{2\,1}(x)\right]\;, \label{hamilt1}
\end{eqnarray}
where
\begin{eqnarray}
U_{j\,k}(x)  =
U_{j}(x)
U_{k}(x+{\hat j}a)
U_{j}(x+{\hat k}a)^{\dagger}
U_{k}(x)^{\dagger} \;,
\nonumber
\end{eqnarray}
and the bare coupling constant $g_{0}$ is dimensionless. Note that the coefficient of the
kinetic term can be written in terms of the continuum coupling constant $e$, namely
$g_{0}^{2}/(2a)=e^{2}/2$. It is for this reason that hadron masses and the 
string tension evaluated
in lattice strong-coupling expansions all scale sensibly with $e$, in $(2+1)$ dimensions.

We denote 
the adjoint representation of the SU($N$) gauge field by $\cal R$:
\begin{eqnarray}
\sum_{c=1}^{N^{2}-1}{\cal R}_{b}^{\;\;c}t_{c}=
Ut_{b}U^{\dagger}\;.
\nonumber
\end{eqnarray}
The matrix ${\cal R}$ lies in the group  
SU($N$)/${\mathbb Z}_{N}$. This
is a special orthogonal matrix ${\cal R}^{\rm T}{\cal R}=1$, 
${\rm det}\;{\cal R}=1$, and 
SU($N$)/${\mathbb Z}_{N}$ is a subgroup
of SO($N^{2}-1$).

Schr\"odinger wave functions are  
complex-valued functions
of {\em all} the link degrees of freedom
$U_{j}(x)$. Physical wave functions $\Psi(\{U\})$ satisfy Gauss' law
\begin{eqnarray}
( {\cal D} \cdot l)(x)_{b}\; \Psi( \{U \}) = \sum_{j=1}^{2} 
\left[ {\cal D}_{j}\; l_{j}(x)\right]_{b}  \;
\Psi(\{U\}) =0 \;,
\label{gauss}
\end{eqnarray}
where 
\begin{eqnarray}
\left[{\cal D}_{j} l_{j}(x)\right]_{b}  = 
l_{j}(x)_{b}-
\sum_{c=1}^{{N}^{2}-1} {\cal R}_{j}(x-{\hat k}a )_{b}^{\;\;\;c} \;l_{j}(x -{\hat k}a )_{c} \;. \label{cov-deriv}
\end{eqnarray}

Sometimes it is useful to introduce color charge operators at lattice sites, denoted by $q(x)_{b}$, which
satisfy
\begin{eqnarray}
[q(x)_{b},q(y)_{c}]={\rm i} f_{bc}^{a} \delta_{xy}q(x)_{a} \; . \label{charge-comm}
\end{eqnarray}
In the presence of charges, Gauss's law becomes
\begin{eqnarray}
\left[( {\cal D} \cdot l)(x)_{b}-q(x)_{b}\right]\; \Psi( \{U \}) =0 \;.
\label{gauss-charge}
\end{eqnarray}

Henceforth, we 
will drop the explicit summation symbol for repeated 
group indices and
adopt the Einstein summation convention.  Sometimes we 
omit the lattice site or link labels, provided no confusion should
be caused by such omissions.

There is a natural geometric interpretation of
the electric-field operator. 
The Maurer-Cartan vector $e_{a}^{\;\;\;a}$, 
on the manifold of SU($N$)
defined by 
\begin{eqnarray}
e_{\alpha}^{\;\;\;a} t_{a}=-i U^{-1}\partial_{\alpha} U\;, \nonumber
\end{eqnarray}
is given explicitly by 
\begin{eqnarray}
e_{\alpha}^{\;\;\;a}  =
-i\left( \frac{ {{\rm 1}\!\!\,{\rm l}} - e^{i{\cal A}\cdot T}}{
{\cal A}\cdot T}\right)_{\alpha}^{\;\;\;a} 
\;,\nonumber
\end{eqnarray}
in canonical coordinates ${\cal A}^{\alpha}$, $\alpha=1,\dots N^{2}-1$, defined by $U=e^{-i{\cal A}\cdot t}$, and 
$\partial_{\alpha}=\partial/\partial {\cal A}^{\alpha}$. The 
coordinates ${\cal A}$ are related to the continuum gauge field $A$ by ${\cal A}=aA$. The matrix
$e$ is nonsingular (including at 
${\cal A}^{\alpha}=0$). One may view 
$e_{a}^{\;\;\;a}$ as the linear map from the group
manifold to the tangent space; this is a particular choice of the 
vielbein, and in this case there is torsion. The electric-field operators are given by
\begin{eqnarray}
l_{a}=-i(e^{-1})_{a}^{\;\;\;\alpha}\partial_{\alpha}\;. 
\nonumber
\end{eqnarray}

\section{The axial gauge on a cylinder}
\setcounter{equation}{0}
\renewcommand{\theequation}{3.\arabic{equation}}

By
fixing an axial gauge,  we will find
the gauge-invariant degrees
of freedom, up to coordinate singularities 
of measure 
zero. Such gauge fixings 
have been discussed many years
ago, both in the continuum  \cite{halpern} and on a 
lattice \cite{halpernbatrouni},  in the path-integral 
approach to gauge theories. The advantage
of working with the Hamiltonian instead of the path 
integral is that unphysical components
of the gauge fields may be more easily eliminated using Gauss's law \cite{ham-ax} (this
could also be done in a transfer matrix formalism).

We
choose space to be a lattice cylinder of size $L^{1}\times L^{2}$, with periodic boundary
conditions in the $2$-direction only. This means that for any function $f(x^{1},x^{2})$
of lattice sites $f(x^{1},x^{2}+L^{2})=f(x^{1}, x^{2})$. We take
components of $x$ to have the values 
$x^{1}=0,a,2a,\dots ,L^{1}$, and $x^{2}=0,a,2a,\dots, L^{2}-a$. Gauss's 
law is still given by (\ref{gauss}), provided
(\ref{cov-deriv}) is modified to
\begin{eqnarray}
{\cal D}_{1} l_{1}(x) & =&  (1-\delta_{x^{1}\;L^{1}})l_{1}(x)
-(1-\delta_{x^{1}\;0}) {\cal R}_{1}(x^{1}-a,x^{2})  l_{1}(x^{1}-a,x^{2})\;,
\nonumber \\
{\cal D}_{2} l_{2}(x) & =& l_{2}(x)
- {\cal R}_{2}(x^{1},x^{2}-a)l_{2}(x^{1},x^{2}-a)\;, \label{cov-deriv1}
\end{eqnarray}
to take into account points on the boundary.

We gauge-fix the links in the $1$-direction by $U_{1}(x^{1}, x^{2})={{\rm 1}\!\!\,l}$ everywhere
and use \rf{gauss} and (\ref{cov-deriv1}) to write
\begin{eqnarray}
l_{1}(x^{1},x^{2}) =  
-\sum_{y^{1}=0}^{x^{1}}( {\cal D}_{2} \cdot l_{2})
(y^{1}, x^{2})
\;. \label{cylgausssoln}
\end{eqnarray}

There is some non-Abelian gauge invariance remaining, namely that
\begin{eqnarray}
\Gamma(x^{2}) \Psi= 
\sum_{x^{1}=0}^{L^{1}}
( {\cal D}_{2} \cdot l_{2})(x^{1},x^{2})\Psi=0
\;. \label{remaining}
\end{eqnarray}

We split the Hamiltonian into two terms $H=H_{0}+\kappa V$, where eventually we
set $\kappa=\frac{g_{0}^{2}}{2a}$
\begin{eqnarray}
H_{0}&\!=\!& \sum_{x^{2}=0}^{L^{2}-a} \left\{
\sum_{x^{1}=0}^{L^{1}} \frac{g_{0}^{2}}{2a}[l_{2}(x^{1}, x^{2})]^{2}  \right. \nonumber \\
&\!-\!&  \left.  \sum_{x^{1}=0}^{L^{1}-a}
\frac{1}{2g_{0}^{2}a}[{\rm Tr}\; U_{2}(x^{1},x^{2})^{\dagger}U_{2}(x^{1}+a,x^{2})+c.c.]
\right\} \;, \label{h-naught}
\end{eqnarray}
and
\begin{eqnarray}
V=\sum_{x^{2}=0}^{L^{2}-a} \sum_{x^{1}=0}^{L^{1}}
\left[ \sum_{y^{1}=0}^{x^{1}}( {\cal D}_{2} \cdot l_{2})(y^{1},x^{2}) \right]^{2}\;.\label{l-one-squared}
\end{eqnarray}

It will be important for the discussion in the next section that the constraint (\ref{remaining})
allows us to replace (\ref{l-one-squared}) by
\begin{eqnarray}
V=-\sum_{x^{2}=0}^{L^{2}-a} \sum_{x^{1}=0}^{L^{1}-a} \left[\sum_{y^{1}=0}^{x^{1}}( {\cal D}_{2} \cdot l_{2})(y^{1},x^{2})\right]^{T}
\left[\sum_{z^{1}=x^{1}+a}^{L^{1}}( {\cal D}_{2} \cdot l_{2})(z^{1},x^{2})\right]
\;.\label{new-V}
\end{eqnarray}

We have assumed until now that no charges are present. If a quark is
placed at site $u$, then (\ref{gauss-charge}) may be solved to give
\begin{eqnarray}
l_{1}(x^{1},x^{2}) =  q(u^{1},u^{2})\delta_{x^{1} \ge u^{1}}\delta_{x^{2}\,u^{2}}
-\sum_{y^{1}=0}^{x^{1}}( {\cal D}_{2} \cdot l_{2})
(y^{1}, x^{2})
\;. \label{cylgausssoln-charge}
\end{eqnarray}
The remaining gauge invariance is
\begin{eqnarray}
\left[ \sum_{x^{1}=0}^{L^{1}}
( {\cal D}_{2} \cdot l_{2})(x^{1},x^{2})- q(u^{1},u^{2} )\delta_{x^{2}\,u^{2}}\right]\Psi=0
\;. \label{remaining-charge}
\end{eqnarray}

\section{Confinement at leading order}
\setcounter{equation}{0}
\renewcommand{\theequation}{4.\arabic{equation}}

The splitting (\ref{h-naught}), (\ref{l-one-squared}) is not $90^{0}$ rotation invariant. Nonetheless, if perturbation theory
in $V$ makes sense, this rotation invariance should be restored at sufficiently high orders. Notice
that $H_{0}$ is a set of decoupled ($1+1$)-dimensional
lattice chiral non-linear sigma models, with global symmetry ${\rm SU}(N)_{L}\times {\rm SU}(N)_{R}$, plus
an extra term at the boundary $x^{1}=0$ (this is a sum of unitary matrix-model Hamiltonians). The on-shell properties of these sigma models have been 
completely
determined; the
Bethe {\em Ansatz} \cite{pol-wieg} and analytic $S$-matrix theory \cite{abda-wieg} determine
the spectrum in the 
renormalized continuum limit.

Let us briefly describe the particles of the chiral ${\rm SU}(N)_{L}\times {\rm SU}(N)_{R}$ model. There are fundamental particles with mass $m_{1}$ transforming as the fully antisymmetric
tensor representation of  ${\rm SU}(N)_{L}\times {\rm SU}(N)_{R}$. The particles are labeled by
a quantum number $n=1,\dots N-1$. The particle with $n>1$
is a bound state of $n$ fundamental particles. We may regard the bound state of $n$
particles as a bound state of $N-n$ antiparticles. There is no singlet
in the one-particle spectrum (which would correspond to $n=N$). The particles have masses
\begin{eqnarray}
m_{n}=m_{1} \frac{\vert \sin\frac{n\pi}{ N}\vert}{ \sin\frac{\pi}{ N} }\;, n=1,\dots, N-1\;,
\nonumber
\end{eqnarray}
where the mass gap $m_{1}$ is of the
form
\begin{eqnarray}
m_{1}=\frac{C}{a} \left(g_{0}^{K_{2}}\;e^{-\frac{K_{1}}{g_{0}^{2}}}
+ \cdots \right)
=   \frac{C}{a} \left[ (e^{2}a)^{\frac{ K_{2}}{2}  }\;\exp{-\frac{K_{1}}{e^{2}a}}
+ \cdots \right]      \;, \label{massgap}
\end{eqnarray}
where $C$ is a non-universal constant, $K_{1}=-1$ and $K_{2}=4\pi$ are determined from the one- and
two-loop coefficients of the chiral-model beta function \cite{mckane}, respectively, and the corrections are non-universal. 

Suppose that there are no charges present. The 
remaining gauge invariance (\ref{remaining}) means that we impose on the states $\Psi$
of the chiral model at $x^{2}$ the constraints
\begin{eqnarray}
\sum_{x^{1}=0}^{L^{1}} R_{2}(x^{1},x^{2}-a) l_{2}(x^{1},x^{2}-a) \Psi=
\sum_{x^{1}=0}^{L^{1}} l_{2}(x^{1},x^{2}) \Psi \;. \label{residual-invariance}
\end{eqnarray}
The meaning of (\ref{residual-invariance}) is that if the state of the
chiral model at some particular $x^{2}$ transforms as a vector with some set of weights
under ${\rm SU}(N)_{L}$,  then the state of the chiral model at $x^{2}+a$ transforms
the same way under ${\rm SU}(N)_{R}$.

For the ground state $\Psi_{0}^{(0)}$, which is a product of chiral model ground states, each side
of (\ref{residual-invariance})  is automatically zero; for 
the Hohenberg-Mermin-Wagner theorem guarantees that it is a singlet under both 
the left global ${\rm SU}(N)_{L}$  and under the right global ${\rm SU}(N)_{R}$ invariances.

The leading-order Hamiltonian describes a theory which confines fundamental charges
separated in the $2$-direction. We will show that this is 
true by two different lines of reasoning. The first proof is more in line with the
way people usually think about phenomena in gauge theories. The second proof
is a direct
utilization of the concepts we have used in the previous section and this one. 

Here is the first proof: suppose that a quark is placed at $u^{1}, u^{2}$ and
an anti-quark at $u^{1}, y^{2}\gg u^{2}$. Gauge-invariant states are of
the form
\begin{eqnarray}
\vert C >= A(u^{1},y^{2})^{\dagger} \prod_{{\rm link}\in C} U({\rm link}) \; B(u^{1}, u^{2} )^{\dagger}\;
\vert 0> \;, \label{meson}
\end{eqnarray}
for some path $C$ of links joing the quark to the anti-quark, whose creation operators
are $A^{\dagger}$ and $B^{\dagger}$, respectively. The lowest-energy state in the presence of the sources is a superposition of such states. The 
Hohenberg-Mermin-Wagner theorem states that for a Hamiltonian with a global
continuous symmetry, there is no spontaneous symmetry breaking. In the unperturbed vacuum, therefore, 
$<0\vert U_{2} \vert 0>=0$. This means that the action of $U_{2}$ on the chiral-sigma-model ground
state produces a superposition of excited states only. Thus the expectation value of $H_{0}$
in any state (\ref{meson}) must be bounded below by the gap times the separation of the
fundamental charges, i.e. 
\begin{eqnarray}
<C\vert H_{0} \vert C>\; \ge \frac{m_{1}}{a}\vert y^{2}-u^{2}\vert  \;, \label{confinement}
\end{eqnarray}
which means that there is confinement of fundamental charges, with string tension 
$m_{1}/a$. We call this phenomenon ``vertical confinement",  because
confinement occurs in the $2$-direction. 

Now for the second proof: the constraint (\ref{remaining-charge}) has the
form
\begin{eqnarray}
\sum_{x^{1}=0}^{L^{1}} R_{2}(x^{1},x^{2}-a) l_{2}(x^{1},x^{2}-a) \Psi
\!&\!-\!&\!
q(u^{1},u^{2})\delta_{u^{2}\,x^{2}} \Psi+q(u^{1},u^{2})\delta_{u^{2}\,y^{2}} \Psi \nonumber \\
\!&\!=\!&\!
\sum_{x^{1}=0}^{L^{1}} l_{2}(x^{1},x^{2}) \Psi \;. \label{residual-invariance-charge}
\end{eqnarray}
This tells us that if the chiral model at $u^{2}-a$ is
in an ${\rm SU}(N)_{L}$ singlet state (such as the vacuum), then the chiral model at $u^{2}$ cannot
be in an ${\rm SU}(N)_{R}$ singlet. Thus the chiral model at $u^{2}$ is in an excited state. By
continuing to use (\ref{residual-invariance-charge}) we conclude that all the chiral models
for $x^{2}$ satisfying $u^{2}\le x^{2}\le y^{2}$ are excited. In this way, we obtain the same result for
the vertical string tension as that given above.

A rectangular Wilson loop of size $S_{1}\times S_{2}$ is
\begin{eqnarray}
A(S_{1}\times S_{2})=  {\rm Tr}\;W(x^{1}, x^{2}; S_{2})^{\dagger}  W(x^{1}+S_{1}, x^{2}; S_{2})
\;, \label{wilson}
\end{eqnarray}
in our gauge, where
\begin{eqnarray}
W(x^{1}, x^{2}; S_{2})=U_{2}(x^{1},x^{2}+S_{2})\cdots U_{2} (x^{1},x^{1}) \;. \nonumber
\end{eqnarray}
Correlation functions of $U_{2}$ decay exponentially. We expect that for large $S_{1}$
\begin{eqnarray}
<0\vert   [U_{2}(x^{1},x^{2})^{\dagger}]_{a}^{\;\;\;b} \;
U_{2}(x^{1}+S_{1},x^{2})_{c}^{\;\;\;d}  \vert0> \;\simeq\; D_{ac}^{bd}\exp\left(-m_{1}S_{1}\right)
\;. \nonumber
\end{eqnarray}
The Wilson loop expectation value is a product of $S_{2}/a$ such
correlation functions, and therefore
\begin{eqnarray}
<0\vert  A(S_{1}\times S_{2}) 
\vert0> \;\simeq \;\exp\left( -\frac{m_{1}}{a}S_{1}S_{2}\right)\;. \label{area-law}
\end{eqnarray}
This is an area law, with the same string tension $m_{1}/a$ found above.

There is no ``horizontal confinement"  -  that is, there is no confinement in the 
$1$-direction - yet.  Horizontal
confinement
will only appear if the perturbation
$\kappa V=g_{0}^{2}V/(2a)$ 
is taken into account.  This is because the constraint consistent with the
presence of 
a quark at
$x^{1}, x^{2}$ and an anti-quark at $y^{1}, x^{2}$ with $y^{1}\gg x^{1}$ is 
(\ref{remaining}), which is satisfied by the unperturbed vacuum. Thus, if 
$V$ is neglected, there is no force
between a 
quark-anti-quark in the $1$-direction. The appearance of 
horizontal confinement in perturbation theory will be demonstrated in Section 6.

\section{Weak-coupling perturbation theory and infrared finiteness}
\setcounter{equation}{0}
\renewcommand{\theequation}{5.\arabic{equation}}

If $L^{1}$ and $L^{2}$ are kept finite, the spectrum of the Hamiltonian $H_{0}+\kappa V$
is purely discrete. Let us consider this spectrum to second order in Rayleigh-Schr\"{o}dinger 
perturbation theory:
\begin{eqnarray}
E_{n}=E_{n}^{(0)}+\kappa E_{n}^{(1)}+\kappa^{2} E_{n}^{(2)}+\cdots \;, \nonumber
\end{eqnarray}
where
\begin{eqnarray}
E_{n}^{(1)}=<\Psi_{n}^{(0)}\vert V\vert \Psi_{n}^{(0)}>,
\;\;E_{n}^{(2)}=-\sum_{m\neq n}
\frac{\vert <\Psi_{n}^{(0)}\vert V\vert \Psi_{m}^{(0)}>\vert^{2}}{E_{m}^{(0)}-E_{n}^{(0)}},\;\dots,
\label{energy-correc}
\end{eqnarray}
and $\vert \Psi_{n}^{(0)}>$ are the eigenvectors of $H_{0}$ with eigenvalues $E_{n}^{(0)}$.  The purpose of this section is to show that the corrections to the vacuum energy $E_{0}^{(1)}$ and $E_{0}^{(2)}$ are proportional
to $L^{1}\times L^{2}$. This happens for two reasons: 1) the ground state of the chiral
model is ``disordered", i.e. two-point functions fall off exponentially, and 2) the 
unperturbed vacuum is a singlet,  simplifies
the form of $V$ acting on this vacuum to (\ref{new-V}).  Our philosophy is 
close to that of Mandelstam \cite{mandelstam} in this regard. 

The first correction to the vacuum energy is
\begin{eqnarray}
E_{0}^{(1)} \!\!&\!=\!&\!\! -<\Psi_{0}^{(0)}\vert  L^{2} \!\sum_{x^{1}=0}^{L^{1}} \sum_{y^{1}=0}^{x^{1}} \sum_{z^{1}=x^{1}+a}^{L^{1}}
[{\mathcal D}_{2}l_{2}(y^{1}, x^{2})]^{T} \nonumber \\
\!\!&\!\times \!&\!\!
{\mathcal D}_{2}l_{2}(y^{1}, x^{2})
\vert \Psi_{0}^{(0)}>
\;.\nonumber
\end{eqnarray}
Correlation functions of $l_{2}$ and $R_{2}l_{2}$ must decay exponentially with the
distance, and therefore this quantity will have the form
\begin{eqnarray}
E_{0}^{(1)} \simeq -L^{2} \sum_{x^{1}=0}^{L^{1}-a} \sum_{y^{1}=0}^{x^{1}} \sum_{z^{1}=x^{1}+a}^{L^{1}}
e^{-m_{1}\vert y^{1}-z^{1}\vert}
\;, \label{first-order-is-okay}
\end{eqnarray}
The dominant contribution to this expression comes from $y^{1}\approx z^{1}$. Since $y^{1}\le
x^{1}<z^{1}$, $E_{0}^{(1)}$ is proportional to the volume $L^{1}L^{2}$. 

Next we sketch the proof that the second-order correction to the vacuum energy also
scales linearly with the volume. Notice that the coefficient of each energy denominator in the second correction (\ref{energy-correc}) 
is non-positive. Thus
\begin{eqnarray}
\vert E_{0}^{(2)} \vert &<&\frac{1}{m_{1}}\sum_{m\neq 0}
\vert <\Psi_{0}^{(0)}\vert V\vert \Psi_{m}^{(0)}>\vert^{2} \nonumber \\
&=&\frac{1}{m_{1}} \left[  <\Psi_{0}^{(0)}\vert V^{2}\vert \Psi_{0}^{(0)}>
-(<\Psi_{0}^{(0)}\vert V\vert \Psi_{0}^{(0)}>)^{2}
\right]
\;. \label{second-order-bound}
\end{eqnarray}
The connected vacuum expectation value on the right-hand-side of (\ref{second-order-bound}) has the following form:
\begin{eqnarray}
\vert E_{0}^{(2)} \vert &<&\frac{L^{2}}{m_{1}}\sum_{x^{1}=0}^{L^{1}}\sum_{w^{1}=0}^{L^{1}}
 C(x^{1},w^{1}; x^{2}) \;, \nonumber
\end{eqnarray}
where
\begin{eqnarray}
\!\!\!\!\!\!C(x^{1},w^{1}; x^{2})\!\!&\!=\!&\!\sum_{{\rm r}} \sum_{y^{1}=0}^{x^{1}}\sum_{z^{1}=x^{1}+a}^{L^{1}} \sum_{u^{1}=0}^{z^{1}}\sum_{v^{1}=w^{1}+a}^{L^{1}}
 \left[  <\Psi_{0}^{(0)}\vert   \right.   \nonumber \\
\!&\!\times\!&\! 
 { \mathcal D}_{2}l_{2}(y^{1},x^{2})^{T}{ \mathcal D}_{2}l_{2}(z^{1},x^{2}) \nonumber \\
\!&\!\times\!&\! { \mathcal D}_{2}l_{2}(u^{1},x^{2}+{\rm r} a)^{T}{ \mathcal D}_{2}l_{2}(v^{1},x^{2}+{\rm r} a) \vert \Psi_{0}^{(0)}>   \nonumber \\
\!&\!-\!&\! 
<\Psi_{0}^{(0)}\vert { \mathcal D}_{2}l_{2}(y^{1},x^{2})^{T}{ \mathcal D}_{2}l_{2}(z^{1},x^{2}) 
\vert   \Psi_{0}^{(0)}  > \nonumber \\
\!&\!\times\!&\! \!\!\!\!     \left.
<\Psi_{0}^{(0)}   \vert 
{ \mathcal D}_{2}l_{2}(u^{1},x^{2}+{\rm r} a)^{T}
{ \mathcal D}_{2}l_{2}(v^{1},x^{2}+{\rm r} a) \vert \Psi_{0}^{(0)}> \right],
\label{second-order-bound1}
\end{eqnarray}
and
where ${\rm r}=0,\pm1$. 
The chiral model is a massive local quantum field theory, so that vacuum correlation functions must cluster for the dominant part of the summations in 
(\ref{second-order-bound1}). Therefore this expression is approximated well by
\begin{eqnarray}
\!C(x^{1},w^{1}; x^{2}) \!&\!\approx\!&\!\sum_{{\rm r}} \sum_{y^{1}=0}^{x^{1}}\sum_{z^{1}=x^{1}+a}^{L^{1}} \sum_{u^{1}=0}^{z^{1}}\sum_{v^{1}=w^{1}+a}^{L^{1}} 
\!\left[  <\!\Psi_{0}^{(0)}\vert { \mathcal D}_{2}l_{2}(y^{1},x^{2})^{T}  \right. \nonumber \\
\!&\!\times\!&\! 
{\mathcal D}_{2}l_{2}(u^{1},x^{2}+{\rm r} a)\!
\vert \Psi_{0}^{(0)} \!\! >
\nonumber \\
\!&\!\times\!&\!  <\Psi_{0}^{(0)}\vert
{ \mathcal D}_{2}l_{2}(z^{1},x^{2})^{T}{ \mathcal D}_{2}l_{2}(v^{1},x^{2}+{\rm r} a)  \vert \Psi_{0}^{(0)}>  
\nonumber \\
\!&\!+\!&\! 
<\Psi_{0}^{(0)}\vert { \mathcal D}_{2}l_{2}(y^{1},x^{2})^{T} { \mathcal D}_{2}l_{2}(v^{1},x^{2}+{\rm r} a)
\vert \Psi_{0}^{(0)}  > \nonumber \\
\!&\!\times\!&\!  \left.<\Psi_{0}^{(0)}\vert
{ \mathcal D}_{2}l_{2}(z^{1},x^{2})^{T}{ \mathcal D}_{2}l_{2}(u^{1},x^{2}+{\rm r} a)  \vert \Psi_{0}^{(0)}>  
\right]
 \;.   \label{clustering}
\end{eqnarray}
By using (\ref{remaining}) we can write each term of (\ref{clustering}) as something which vanishes
exponentially away from $x^{1}=w^{2}$. For example, consider the first factor of the first
term:
\begin{eqnarray}
{\rm First\; Factor}=\sum_{y^{1}=0}^{x^{1}} \sum_{u^{1}=0}^{z^{1}}
<\Psi_{0}^{(0)}\vert { \mathcal D}_{2}l_{2}(y^{1},x^{2})^{T}
{ \mathcal D}_{2}l_{2}(u^{1},x^{2}+{\rm r} a)\vert \Psi_{0}^{(0)}  > .
\label{first-factor}
\end{eqnarray}
If $x^{1}\le z^{1}$, we may write this as
\begin{eqnarray}
{\rm F.F.}=-\sum_{y^{1}=0}^{x^{1}} \sum_{u^{1}=z^{1}+a}^{L^{1}}
<\Psi_{0}^{(0)}\vert { \mathcal D}_{2}l_{2}(y^{1},x^{2})^{T}
{ \mathcal D}_{2}l_{2}(u^{1},x^{2}+{\rm r} a)\vert \Psi_{0}^{(0)} \! > ,
\label{first-factor1}
\end{eqnarray}
and we see that this expression is finite as $L^{1}\rightarrow \infty$. On the other
hand, if $x^{1}\ge z^{1}$, we rewrite (\ref{first-factor}) as
\begin{eqnarray}
{\rm F.F.}=-\sum_{y^{1}=x^{1}+a}^{L^{1}} \sum_{u^{1}=0}^{z^{1}}
<\Psi_{0}^{(0)}\vert { \mathcal D}_{2}l_{2}(y^{1},x^{2})^{T}
{ \mathcal D}_{2}l_{2}(u^{1},x^{2}+{\rm r} a)\vert \Psi_{0}^{(0)} \! > ,
\label{first-factor2}
\end{eqnarray}
and reach the same conclusion. Since each factor of each term behaves this way, we
can conclude that the second-order correction to the vacuum energy can increase at most
linearly with $L^{1}$.

Infrared finiteness of the vacuum energy to first and second order  
in perturbation theory inspires
confidence that it should hold to all orders. The main complication beyond the second
order is the lack of
non-positivity or non-negativity of products of matrix elements. We believe that careful
application of the linked-cluster expansion, assuming clustering in the chiral sigma
model, can provide a proof to all orders.

\section{Horizontal confinement}
\setcounter{equation}{0}
\renewcommand{\theequation}{6.\arabic{equation}}

In Section 4 we showed that quarks are confined vertically, in the $2$-direction, but
not horizontally, in the $1$-direction, at the zeroth order of the weak-coupling expansion. To
see what happens beyond this order, it is necessary to examine the quark-anti-quark
potential in perturbation theory. This is very straightforward to do.

If a quark is located at $u^{1},u^{2}$, and an anti-quark is located at $v^{1},u^{2}$ with
$v^{1}>u^{1}$, the electric-field operator in the $1$-direction is given by
\begin{eqnarray}
l_{1}(x^{1},x^{2}) =  q(u^{1},u^{2})\delta_{x^{1} \ge u^{1}}\delta_{x^{2}\,u^{2}}
\!&\!-\!&\!q(v^{1}, u^{2})\delta_{x^{1} \ge v^{1}}\delta_{x^{2}\,u^{2}} \nonumber \\
\!&\!-\!&\!\sum_{y^{1}=0}^{x^{1}}( {\cal D}_{2} \cdot l_{2})
(y^{1}, x^{2})
\;. \label{two-quarks}
\end{eqnarray}
The constraint (\ref{remaining}) is unmodified. Thus, the unperturbed states
and energies are unaffected by these two charges, as we claimed in Section 4. However, to first 
order in perturbation theory, there is a new contribution to $E_{0}^{(1)}$ equal to
\begin{eqnarray}
\Delta E_{0}^{(1)}= \kappa C_{N} \vert v^{1}-u^{1} \vert\;, \label{horizontal-conf}
\end{eqnarray}
where $C_{N}$ is the smallest eigenvalue of the Casimir of 
${\rm SU}(N)$, $q^{2}=C_{N}{\rm 1}\!\!\,{\rm l}$, by (\ref{charge-comm}). Thus, to first
order in perturbation theory, the horizontal string tension is 
$\frac{\kappa}{a}C_{N}$.  What is 
especially remarkable about this result is 
that we can see clearly an electric string forming along the shortest path
connecting the two quarks. 

What is happening physically is that the vacuum remains undisturbed by the charges
and prevents the penetration of electric flux. To this low order of perturbation theory, we
have a cost of at least $m_{1}$ to excite the chiral model at $x^{2}$ and $x^{2}-a$. Thus 
there is a string tension equal to the $(1+1)$-dimensional string tension
through an electric Meissner effect. We do not
have to appeal
to the condensation of some kind of magnetic charge
to make this interpretation.  At higher orders of perturbation theory, the string of electric
flux can presumably fluctuate; these corrections are needed to reliably
set  $\kappa=\frac{g_{0}^{2}}{2a}$.

\section{Conclusions}
\setcounter{equation}{0}
\renewcommand{\theequation}{7.\arabic{equation}}

In this paper, we have shown that lattice gauge theories in two space and one time dimension
confine charges, through an anisotropic weak-coupling expansion. Though we cannot
exactly evaluate the terms in this expansion, by just using some general knowledge of the
chiral nonlinear sigma models, we can make precise statements about these terms.

The astute reader may wonder if the methods developed here can work for the oldest
known example of non-trivial confinement: lattice
compact $({\rm QED})_{2+1}$ \cite{polyakov1}. The answer is that they do not. In this Abelian gauge
theory, we would expand about the states of the ${\rm U}(1)$ nonlinear sigma model. This
model has a massless phase at weak coupling, so we would not obtain vertical confinement
and area-law behavior of the space-like Wilson loop. In fact, our
perturbation method
makes no sense at all for lattice $({\rm QED})_{2+1}$. The reason is that correlations of the
operator $l(x^{1},x^{2})-l(x^{1},x^{2}-a)$ (the adjoint-representation covariant derivative
is simply the ordinary lattice derivative) do not fall off sufficiently fast to make
$\sum l_{1}^{2}$ directly 
proportional to the volume. The infrared divergence in the vacuum energy,
which concerned us so 
much, really happens in the Abelian theory. This divergence is not real, but is an artifact of our
methods. Our weak-coupling expansion seems peculiarly 
suited to non-Abelian theories
in this regard. 

We have assumed that a mass gap exists in the $(1+1)$-dimensional ${SU}(N)$ chiral
model. At strong coupling, this can be proved rigorously with a cluster expansion,
in the Euclidean lattice formulation. Perhaps a fully rigorous proof can be made of
confinement with $g_{0}$ large, but $\kappa$ small. 

Of the questions raised by our analysis we think that six
stand out as important.  We suspect, however,  that only the 
first, second  and possibly the third can be answered in the near future.

The first and probably easiest important question is whether the infrared finiteness of our
perturbation series exists beyond the second order. We hope to be able to settle
this issue soon. If settled affirmatively, the
series probably does not converge, but may be Borel summable.

The second question is whether adjoint matter is confined for finite $N$. This is certainly happening
at first order in the horizontal direction. We believe that this property will disappear at higher
orders.

The third question is raised by the fact that our mass scales are set by 
(\ref{massgap}), with one exception (the horizontal string tension). All these quantities are
non-zero for any positive value of $a$. We believe, however, that we should still
have a mass gap and gap confinement as $a\rightarrow 0$, provided the continuum
coupling constant $g_{0}/({\sqrt a})$ is kept fixed. Our vertical string tension, found in Section 4 is too small, and our horizontal string
tension found in Section 6 is known only for small $\kappa$. These 
numbers should both be proportional to 
$g_{0}^{2}/a$, the square of the continuum
coupling constant. Perhaps this difficulty can be 
removed by resummation of the perturbation series or 
by a renormalization-group argument.

The fourth question is whether
we can do a better job of calculating energies and states. Perhaps we could accomplish 
this, if Bethe's {\em Ansatz} for the chiral model could be 
carried out in a formalism where both the left- and right-handed ${\rm SU}(N)$ symmetries
are manifest in the Hamiltonian. In the work of Polyakov and Wiegmann \cite{pol-wieg}
only one of these
is manifest; the other appears in the $S$-matrix, but its interpretation is obscure. If a version
of Bethe's {\em Ansatz} with both symmetries manifest
can be found, there is the possibility of a better
understanding of the $(2+1)$-dimensional gauge theory. One could use whatever regularization
is most expedient for diagonalising the Hamiltonian, instead 
of the lattice. It may be a long time before this question can be seriously
addressed (perhaps never). We believe a more likely path to success is to expand some version
of the axial-gauge
Hamiltonian about a system of $(1+1)$-dimensional field theories other than chiral 
sigma models. It would be a stroke of good luck, to have an expansion about exactly solvable field theories where the
symmetries are easy to understand.

The fifth question is whether our results can be understood in the context of
condensation of magnetic charge. If a picture of condensing composite operators could work in
the $(1+1)$-dimensional chiral models (no one has succeeded in showing this), then
operators defined on sets of points of one dimension higher should be important for confinement in
$(2+1)$ dimensions.

The last and most important question is whether $({\rm QCD})_{3+1}$ could be 
studied by our methods. This is, we hope to no one's 
surprise, a much harder problem. A lattice gauge theory in $(2+1)$ dimensions
is particularly amenable to the methods discussed here, because if the square of
electric field in the $1$-direction is dropped from the Hamiltonian, it 
easily breaks apart into $(1+1)$-dimensional Hamiltonians we know a lot
about. This does not happen in
$3+1$ dimensions. The Hamiltonian breaks into $(2+1)$-dimensional Hamiltonians with
both gauge fields and matter in the adjoint 
representation. These models are probably not even renormalizable, but seem worthy
of investigation.

\section*{Acknowledgements}

I thank Mike Creutz, John Kogut, Herbert Neuberger, and Tilo Wettig for
organizing the workshop ``Modern Challenges for Lattice
Field Theory", at the Kavli Institute for Theoretical Physics, where this work was completed amid
the discussion of many interesting topics. I would also like to thank David Adams for discussions
and a careful reading of the manuscript.

This research was supported in
part by the National Science Foundation under Grant No. PHY99-07949. It
was also supported in
part by a grant from the PSC-CUNY.

\end{document}